# MIMO IDENTICAL EIGENMODE TRANSMISSION SYSTEM (IETS) – A CHANNEL DECOMPOSITION PERSPECTIVE

*M. Zeeshan Shakir[1], Student member IEEE, and Tariq S. Durrani[2], Fellow IEEE*

Department of Electronic and Electrical Engineering, University of Strathclyde
204 George Street, G1 1XW, Glasgow, UK
phone: + (44) 0141 548 2686, fax: + (44) 0141 552 2487, email: {mshakir[1], tsd[2]}@eee.strath.ac.uk
web: www.eee.strath.ac.uk

## ABSTRACT

*In the past few years considerable attention has been given to the design of Multiple-Input Multiple-Output (MIMO) Eigenmode Transmission Systems (EMTS). This paper presents an in-depth analysis of a new MIMO eigenmode transmission strategy. The non-linear decomposition technique called Geometric Mean Decomposition (GMD) is employed for the formation of eigenmodes over MIMO flat-fading channel. Exploiting GMD technique, identical, parallel and independent transmission pipes are created for data transmission at higher rate. The system based on such decomposition technique is referred to as MIMO Identical Eigenmode Transmission System (IETS). The comparative analysis of the MIMO transceiver design exploiting non-linear and linear decomposition techniques for variable constellation is presented in this paper. The new transmission strategy is tested in combination with the Vertical Bell Labs Layered Space Time (V-BLAST) decoding scheme using different number of antennas on both sides of the communication link. The analysis is supported by various simulation results.*

## 1. INTRODUCTION

Exploiting multiple antennas at the transmitter and the receiver in wireless communication systems is an extremely promising way to enhance the data rate of future wireless communication systems [1] and [2]. Multiple-Input Multiple-Output (MIMO) communication system that transmit data through parallel sub-channels and exploit the Channel State Information (CSI) at the transmitter, are termed as MIMO Eigenmode Transmission System (EMTS). However, in such systems, rich-scattering and flat-fading conditions are a requisite condition, in order to exploit high MIMO EMTS capacity [3].

By deploying multiple transmitters and receivers in sufficiently rich in scattering environment, the MIMO EMTS has a high potential to increase the capacity linearly with the number of spatial channels [4]. The same has been illustrated in [4] and [5] for the Bell Labs Layered Space Time (BLAST) architecture. Since then, many transmission strategies have been proposed [1] and [2]. Among the two most important and well known approaches, one is space time coding method that attempts to improve the communication reliabilies by coding and diversity gain achieved by appropriate coding design [1]. The other is spatial multiplexing method which transmits data over spatial sub-channels, often in conjunction with an outer channel code, e.g., the BLAST architecture which focuses on maximizing the channel throughput [1] and [6]. Both design schemes assume that the CSI is available only at the receiver. By transmitting through parallel, spatial sub-channels and exploiting the CSI at the receiver, spatial multiplexing systems can provide high data rates [3]. However, if the communication environment is slowly time varying, such as indoor communication via Wireless Local Area Networks (WLANs), the availability of the CSI is also possible at the transmitter, using feedback or the reciprocal technique when Time Division Duplex (TDD) is used [3]. With the CSI available at the transmitter as well, channel capacity can be achieved exploiting a linear transformation at the transmitter. The linear decomposition techniques can be employed at the receiver to convert the MIMO EMTS channel into set of parallel and independent scalar sub-channels.

The design of MIMO EMTS transceiver includes precoing at the transmitter and employing an equalizer at the receiver [1]. Several designs have been proposed based on conventional decomposition techniques and using a variety of criteria, including maximum Signal to Noise Ratio (SNR), Minimum Mean Squared Error (MMSE) and Bit Error Rate (BER) based criteria [1] and [7].

In [1], a transmission strategy has been designed using linear transformation only. The linear transformation used is Singular Value Decomposition (SVD), which decomposes the MIMO EMTS flat-fading channel into multiple parallel sub-channels. The water filling algorithm is used to achieve the capacity of each sub-channel [1]. The non-zero singular values of the diagonal matrix represent the Signal to Noise Ratios SNRs) of the sub-channels formed by SVD. However, due to very high variations in the SNRs of the sub-channels and high condition number, this apparently simple linear decomposition scheme requires a very intelligent and adaptive bit allocation in order to match the capacity of each sub-channel and achieve the prescribed BER [1]. Bit allocation among the eigenmodes of the MIMO EMTS not only increases the coding/decoding complexity but also inherently capacity loss because of finite constellation granularity [1] and [4]. Alternatively, assignment of equal power to the sub-channels, results in the same constellation among all eigenmodes. However, more transmitting power could be allocated to the channel having a poorer SNR i.e. the eigenmode having lowest SNR. A fundamental trade off is always required to be made between capacity and BER performance when dealing with linear decomposition technique.

In this paper, we first review linear decomposition technique referred to as SVD. Then we introduce non-linear decomposition technique called Geometric Mean Decomposition (GMD). The main aim of this paper is to develop a MIMO EMTS transceiver design exploiting GMD in combination with V-BLAST decoding. Using GMD, the MIMO EMTS channel can be decomposed into set of identical, parallel and independent sub-channels. This non-linear decomposition technique brings about much more convenience in coding/decoding and modulation/demodulation processes. In order to demonstrate the effectiveness of new transmission strategy, various simulation scenarios are created in Matlab. The BER performance curves are obtained for different number of antennas at the transmitter and the receiver with two different types of modulation schemes. Results of the transmission strategy based on non-linear decomposition technique ensure superior performance as compare to conventional transmission strategy based on SVD [1].

The remainder of the paper is organized as follows. Section 2 introduces channel model and decomposition techniques. This section presents an overview of channel decomposition perspective of Rayleigh flat-fading MIMO channel and associated challenges. After reviewing the well understood, linear decomposition technique SVD, we introduce GMD to construct the eigenmode over MIMO channel. This section also presents mathematical concept for the formation of identical, parallel and independent sub-channels. Section 3 presents the testing, comparative results and analyses of GMD based V-BLAST detection scheme.

The following notation is used throughout this paper. Boldface uppercase letters denote matrices, boldface italic lowercase letters denote column vectors, and italics denote scalars. $\mathbb{R}^{M \times N}$ and $\mathbb{C}^{M \times N}$ represent the set of $M \times N$ matrices with real and complex valued entries respectively. The superscripts $(.)^T$ and $(.)^H$ denote transpose and hermitian operation respectively.

## 2. CHANNEL MODEL AND DECOMPOSITION TECHNIQUES

In this section of paper, we present flat-fading channel model and GMD technique to design the precoder at the transmitter and the equalizer at the receiver. We also discuss the formation of identical, parallel and independent transmission pipes over MIMO channel.

### 2.1 The Channel Model

Let us consider a MIMO EMTS having $N$ transmitting antennas and $M$ receiving antennas as shown in Figure 1.

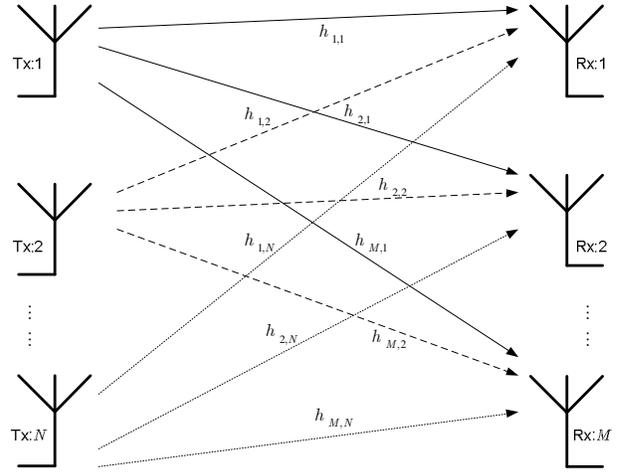

Figure 1 Model of MIMO communication system with $N$ transmitting and $M$ receiving antennas over Rayleigh flat-fading channel.

The signal model corresponding to a transmission through a flat-fading MIMO channel is given by [1] and [3]
$$y = \mathbf{H}s + n \quad (1)$$
where $s \in \mathbb{C}^{N \times 1}$ is the transmitted vector, $\mathbf{H} \in \mathbb{C}^{M \times N}$ is the channel matrix with the $(m,n)th$ element denoting the fading coefficient between the $mth$ receiving and $nth$ transmitting antennas, $y \in \mathbb{C}^{M \times 1}$ is the received signal vector, and $n \in \mathbb{C}^{N \times 1}$ is a zero mean circularly symmetric complex Gaussian interference-plus-noise vector with arbitrary covariance matrix. Throughout this paper we assumed that 1) $K$ denote the rank of $\mathbf{H}$ such that $K \leq \min(M,N)$ and 2) perfect CSI at the transmitter and the receiver i.e. $\mathbf{H}$ is known at both sides of the communication link.

### 2.2 Geometric Mean Decomposition Algorithm

In order to investigate the geometric mean decomposition technique, we review linear transformation of the channel matrix $\mathbf{H}$ called SVD and is given by
$$\mathbf{H} = \mathbf{U} \boldsymbol{\Sigma} \mathbf{V}^H . \quad (2)$$
where $\mathbf{U}$ and $\mathbf{V}$ are unitary matrices and $\boldsymbol{\Sigma}$ is a diagonal matrix with singular values equal to the square root of eigenvalues of the Wishart matrix, given by $\sqrt{\lambda_1}, \cdots, \sqrt{\lambda_K}$ such that following condition is satisfied
$$\sqrt{\lambda_1} \geq \sqrt{\lambda_2} \geq \cdots \geq \sqrt{\lambda_K} > 0 . \quad (3)$$
By calculating the filtered receive vector, we have
$$r = \mathbf{U}^H y , \quad (4)$$
By substituting (1) and apply decomposition defined in (2), we have
$$= \mathbf{U}^H \mathbf{H}s + \mathbf{U}^H n = \mathbf{U}^H \mathbf{U} \boldsymbol{\Sigma} \mathbf{V}^H s + \mathbf{U}^H n ,$$
$$r = \boldsymbol{\Sigma} s' + n' , \quad (5)$$
where $s' = \mathbf{V}^H s$ and $n' = \mathbf{U}^H n$.

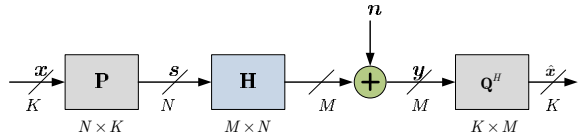

Figure 2 Scheme of MIMO communication system with non-linear transceiver over Rayleigh flat-fading channel.

The MIMO EMTS defined in (1) is now decompose into parallel sub-channels, each representing single in-put single out put (SISO) system given by

$$r_i = \sqrt{\lambda_i}\, s'_i + n'_i. \quad (6)$$

where $i = 1, 2, \cdots, k, \cdots, K$ and $\lambda_i$ denotes the eigenvalues of wishart matrix. It is obvious that the information can only be transmitted over those equivalent channels with non zero singular values [1]. It is very important to note that if the number of transmission layers exceeds the number of strong singular values, the performance of the MIMO system degrades. The complexity in the design of linear transceiver is due to large variations among the eigenvalues of the channel matrix $\mathbf{H}$ [1] and [3].

For any rank $K$, matrix $\mathbf{H} \in \mathbb{C}^{M \times N}$ with singular values $\sqrt{\lambda_1} \geq \sqrt{\lambda_2} \geq \cdots \geq \sqrt{\lambda_K} > 0$, there exist an upper triangular matrix $\mathbf{R} \in \mathbb{R}^{K \times K}$ with identical diagonal elements given by [7] and [8]

$$r_{ii} = \bar{\lambda} = \left(\prod_{i=1}^{K} \lambda_i\right)^{1/2K}. \quad 1 \leq i \leq K \quad (7)$$

such that the singular value decomposition of $\mathbf{R}$ is given by

$$\mathbf{R} = \mathbf{U}_R \mathbf{\Sigma} \mathbf{V}_R^H, \quad (8)$$

with $ith$ diagonal element equal to $r_{ii} = \bar{\lambda}$. Here $\mathbf{\Sigma}$ is a diagonal matrix whose elements are equal to the singular values of the matrix $\mathbf{H}$ i.e. $\{\sqrt{\lambda}\}_{i=1}^{K}$. Since we also know that $\mathbf{H} = \mathbf{U}\mathbf{\Sigma}\mathbf{V}^H$ and combining with (8), we have

$$\mathbf{H} = \mathbf{U}\mathbf{U}_R^H \mathbf{R} \mathbf{V}_R \mathbf{V}^H, \quad (9)$$

Hence, the decomposition derived in (9), is referred to as the Geometric Mean Decomposition (GMD) [3] and [8]

$$\mathbf{H} = \mathbf{Q}\mathbf{R}\mathbf{P}^H. \quad (10)$$

where $\mathbf{Q}$ and $\mathbf{P}$ are semi-unitary matrices denoting the linear operations at the receiver and the transmitter respectively.

### 2.3 Transmission over Identical and Parallel Pipes

The scheme of a general MIMO EMTS communication system with non-linear transceiver is shown in Figure 2. First we calculate the GMD of channel matrix $\mathbf{H}$ as $\mathbf{H} = \mathbf{Q}\mathbf{R}\mathbf{P}^H$. Next we encode the information symbols $\mathbf{s}$ via the linear precoder $\mathbf{P}$, such that the transmitted vector is given by

$$\mathbf{s} = \mathbf{P}\mathbf{x}. \quad (11)$$

where $\mathbf{P} \in \mathbb{C}^{N \times K}$ is the transmit matrix (precoder) and $\mathbf{x} \in \mathbb{C}^{K \times 1}$ is the data vector that contains the $K$ symbols to be transmitted (zero mean, normalized and uncorrelated, that is, $\mathrm{E}[\mathbf{x}\mathbf{x}^H] = \mathbf{I}$) drawn from a set of constellations.

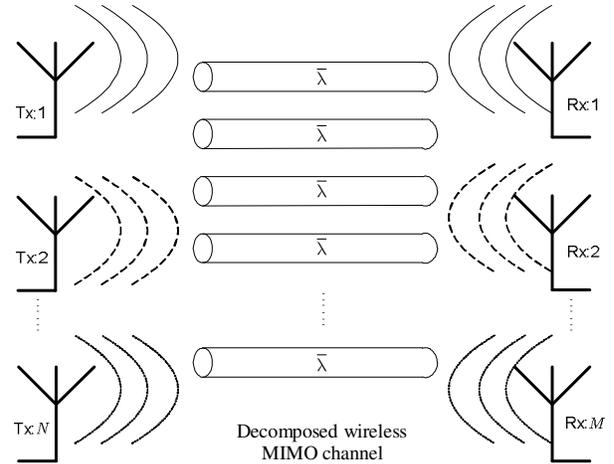

Figure 3 Concept of formation of identical, parallel and independent pipes over Rayleigh flat-fading MIMO channel using Geometric Mean Decomposition (GMD).

At the receiver side, the decomposition algorithm is exploited in combination with a receiver interface, referred to as V-BLAST decoding. The decoding algorithm is based on sequential nulling and cancellation in order to decode the transmitted information symbols $\mathbf{s}$ [1], [6] and [7]. The nulling step can be implemented by either using zero forcing (ZF) or minimum mean squared error (MMSE) criterion [1]. The main drawback of the V-BLAST detection algorithm lies in the computational complexity, as it requires multiple calculations of the pseudoinverse of channel matrix in the ZF case [7]. Thus, we consider GMD scheme in order to design a reduced complexity version of V-BLAST detection scheme.

The estimated data vector at the receiver is given by

$$\hat{\mathbf{x}} = \mathbf{Q}^H \mathbf{y}, \quad (12)$$

where $\mathbf{Q} \in \mathbb{C}^{K \times M}$ is the receiver matrix (equalizer). Substituting (1) in (12), we have

$$\hat{\mathbf{x}} = \mathbf{Q}^H \mathbf{H}\mathbf{s} + \mathbf{Q}^H \mathbf{n}, \quad (13)$$

we consider the scheme presented in [7]. Restating the successive interference cancellation scheme employing GMD (10), the resulting received vector (13) becomes

$$\hat{\mathbf{x}} = \mathbf{Q}^H \mathbf{Q}\mathbf{R}\mathbf{P}^H \mathbf{s} + \mathbf{z}, \quad (14)$$

also substituting (11) in above, we have

$$\hat{\mathbf{x}} = \mathbf{Q}^H \mathbf{Q}\mathbf{R}\mathbf{P}^H \mathbf{P}\mathbf{x} + \mathbf{z}, \quad (15)$$

knowing $\mathbf{Q}\mathbf{Q}^H = \mathbf{I}$ & $\mathbf{P}\mathbf{P}^H = \mathbf{I}$, (15) becomes

$$\hat{\mathbf{x}} = \mathbf{R}\mathbf{x} + \mathbf{z}. \quad (16)$$

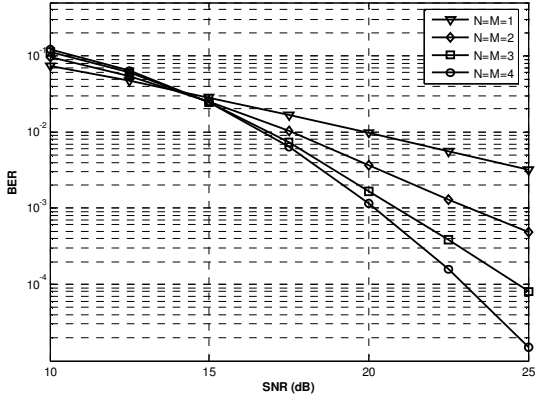

(a)

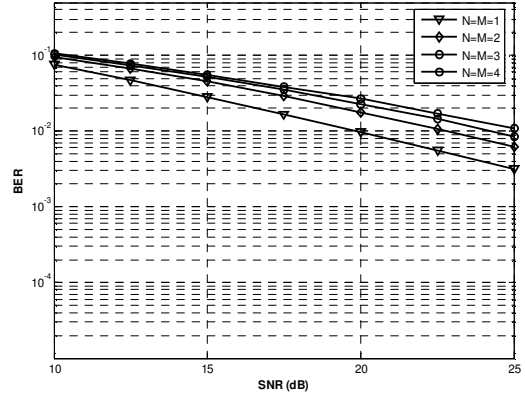

(a)

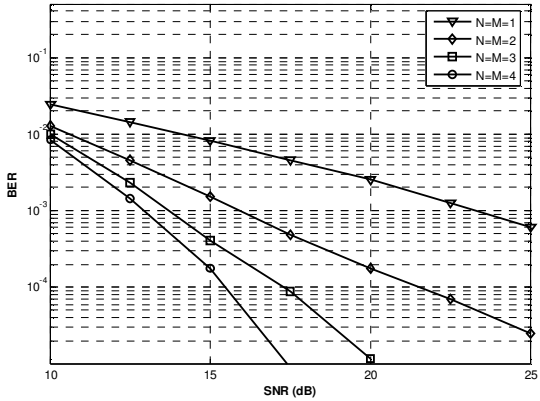

(b)

Figure 4 BER performance of GMD V-BLAST over Rayleigh flat-fading MIMO IETS with (a) QAM-16 (b) QPSK.

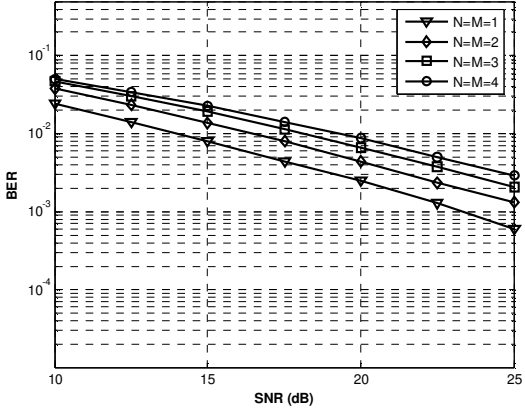

(b)

Figure 5 BER performance of SVD V-BLAST over Rayleigh flat-fading MIMO system with (a) QAM-16 (b) QPSK.

also in component wise notation, (16) becomes

$$\begin{bmatrix} \hat{x}_1 \\ \hat{x}_2 \\ \vdots \\ \hat{x}_K \end{bmatrix} = \begin{bmatrix} r_{1,1} & r_{1,2} & \cdots & r_{1,K} \\ \vdots & r_{2,2} & & \vdots \\ \vdots & & \ddots & \vdots \\ \mathbf{0} & \cdots & \cdots & r_{K,K} \end{bmatrix} \begin{bmatrix} x_1 \\ x_2 \\ \vdots \\ x_K \end{bmatrix} + \begin{bmatrix} z_1 \\ z_2 \\ \vdots \\ z_K \end{bmatrix}. \quad (17)$$

Due to upper triangular structure of **R**, the $ith$ element of $\hat{\boldsymbol{x}}$ is given by

$$\hat{x}_i = r_{i,i} x_i + \sum_{i+1}^{K} r_{i+1,i} x_{i+1} + z_i, \quad (18)$$

But for GMD, we know that $r_{ii} = \overline{\lambda}$ for $i=1,2,\cdots,k,\cdots,K$. Ignoring propagation effects i.e. $\sum_{i+1}^{K} r_{i+1,i} x_{i+1} = 0$, we can regard the resulting sub-channels as $K$ identical, parallel and independent sub channels given by

$$\hat{x}_i = \overline{\lambda} x_i + z_i. \quad for\ i=1,\cdots,K \quad (19)$$

The concept of formation of identical, parallel and independent pipes using GMD is shown in Figure 3. The main advantage of this combined strategy comes with the complexity reduction, as it requires only a fraction of computational effort as compare to the original V-BLAST algorithm [1], [2] and [7]. The channel gain of each eigenmode is given by $\overline{\lambda}$. Beam steering techniques on both the transmitter and the receiver sides are achieved by multiplying vector $\boldsymbol{p}_m$ at the transmitter and matrix $\mathbf{Q}^H$ at the receiver, where $\boldsymbol{p}_m$ denotes $mth$ column of $\mathbf{P}$. As a result, an equivalent channel matrix can be expressed as

$$\mathbf{Q}^H \mathbf{H} \boldsymbol{p}_m = \begin{bmatrix} 0, \cdots, \overline{\lambda}, \cdots, 0 \end{bmatrix}^T. \quad (20)$$

In MIMO systems, data transmission over the equivalent channel given by (20) is referred to as identical eigenmode transmission system (IETS) and the sub-channels and $\boldsymbol{p}_m$ are called as eigenmodes and eigenvectors, respectively.

## 3. PERFORMANCE ANALYSIS

In this Section, we present the performance analysis based on BER curves obtained after various simulation scenarios. In all simulation experiments, we assumed that the channel is Rayleigh flat-fading. To determine the effectiveness of the new GMD based V-BALST detection strategy, the BER performance curves are compared with SVD based V-BLAST decoding strategy. In each scenario the curves are obtained after averaging 5000 Monte Carlo trails of $\mathbf{H}$.

In first case, we consider GMD V-BLAST over Rayleigh flat-fading MIMO channel with $N = 1, 2, 3, 4$ and $M = 1, 2, 3, 4$ transmitting and receiving antennas respectively. In Figure 4(a), we present a scenario where $N$ independent symbols, modulated as Quadrature Amplitude Modulation (QAM-16) are transmitted. Using GMD over $\mathbf{H}$, we have $K$ identical, parallel and independent sub-channels for symbol transmission. After observing the BER curves for different number of transmitting and receiving antennas it is demonstrated that for the curve $N = M = 4$, GMD V-BLAST performs superior, from moderate to high SNRs. Hence, the performance of GMD based V-BLAST scheme is increased with the increase in the number of transmitting and receiving antennas. With the increase in channel dimension the condition number is usually high as well and therefore, the performance curves for channel with higher condition number outperforms the curves with lower condition number. In Figure 4(b), the same simulation scenario is repeated for a different constellation. In this simulation $N$ independent symbols, modulated as Quaternary Phase Shift Keying (QPSK) are transmitted over identical, parallel and independent sub-channels. It is observed that the BER performance curves for QPSK outperform the curves obtained for QAM-16.

Next, we present the simulation scenarios for MIMO linear transceiver over Rayleigh flat-fading channel again with $N = 1, 2, 3, 4$ and $M = 1, 2, 3, 4$ transmitting and receiving antennas respectively. In Figure 5(a), we present a scenario where $N$ independent symbols, modulated as QAM-16 are transmitted. Exploiting SVD over $\mathbf{H}$, we have $K$ parallel and independent sub-channels for symbol transmission. From the BER performance curves it is observed that with the increase in number of transmitting and receiving antennas the performance of the SVD based V-BLAST detection is degraded. For higher channel dimension i.e. $\mathbf{H} \in \mathbb{C}^{4 \times 4}$ the condition number of $\mathbf{H}$ is usually very high that results in ill conditioned sub-channels. Allocating more power to the poor channel degrades the BER performance of the system. To avoid the system degradation, SVD should be used in combination with water filling algorithm which suggests that the binary phase shift keying (BPSK) or QPSK should be used to match the capacity of the worst sub-channels and something like QAM-16 or QAM-64 to the best sub-channels, the same is observed in following simulation scenario. In Figure 5(b), the scheme is tested for same number of transmitting and receiving antennas. The $N$ independent symbols, now modulated as QPSK i.e. same constellation over all eigenmodes, are transmitted over parallel and independent sub-channels. The BER performance curves show better performance as compared to the curves obtained with QAM-16 using SVD based V-BLAST detection scheme.

## 4. CONCLUSION

Efficient and less complex non-linear decomposition technique referred to as GMD is used for the formation of identical, parallel and independent sub-channels over Rayleigh flat-fading MIMO channel. The comparison of GMD and SVD based V-BLAST detection scheme using BER performance curves is being done. The effectiveness of the strategy is observed for various numbers of transmitting and receiving antennas. In SVD based transceiver design we cannot use the same constellation among all parallel sub-channels. Therefore, it involves the use of adaptive bit allocation to match the capacity of each sub-channel which increases the complexity. A trade off is always required to be defined between the capacity and the BER when dealing with SVD based strategy. On the other hand, for GMD technique, the same constellation with moderate size e.g. QAM-16 or QAM-64 can be used to match the capacity of most of the sub-channels. Results of new MIMO IETS ensure superior performance due to BER and complexity in comparison with conventional MIMO transmission system.

## ACKNOWLEDGEMENT

The authors would like to acknowledge the financial support of Picsel Technologies Ltd, Glasgow, UK.